\documentclass{article}
\usepackage{tikz}
\usepackage{graphicx}
\usepackage[utf8]{inputenc}
\usepackage{mathtools}
\usepackage{amsmath}
\usepackage{pgfplots}
\usepackage{breqn}
\DeclarePairedDelimiter\ceil{\lceil}{\rceil}

\usepackage{subfiles}
\usepackage[justification=centering,font=scriptsize,skip=0pt, belowskip=-1pt,aboveskip=0pt]{caption}
\usepackage{amsmath}
\usepackage{epstopdf}
\usepackage{lipsum}
\usepackage[noadjust]{cite}
\usepackage[ruled, vlined, linesnumbered]{algorithm2e}
\SetAlFnt{\small\sffamily}

\SetCommentSty{mycommfont}
\usepackage{color}
\usepackage{pifont}
\usepackage{titlesec}

\DeclareMathOperator*{\argmax}{argmax}
\DeclareMathOperator*{\argmin}{argmin}

\newcommand{\systemCapacity}{n\xspace}

\newcommand{\churnStabilization}{Interlaced\xspace}
\newcommand{\bruijnStateSize}{x\xspace}

\newcommand{\backupUpdate}{backupUpdate}
\newcommand{\backupResolve}{backupResolve}
\newcommand{\swdbg}{SW-DBG\xspace}

\newcommand{\backupSize}{b\xspace}
\newcommand{\stateSet}{\mathcal{S}_{\bruijnStateSize}\xspace}

\newcommand{\predictionGain}{1.11\xspace}
\newcommand{\searchTimeGain}{2.47\xspace}
\newcommand{\successRatioGain}{1.81\xspace}
\newcommand{\uniformFailureProb}{q\xspace}
\newcommand*\xor{\mathbin{\oplus}}

\title{Interlaced: Fully decentralized churn stabilization \\ for Skip Graph-based DHTs}
\author{Yahya Hassanzadeh-Nazarabadi, Alptekin Küpçü and Öznur Özkasap\\Department of Computer Engineering, Koç University, İstanbul, Turkey\\
{\{yhassanzadeh13, akupcu, oozkasap\}}@ku.edu.tr}
\begin{document}

{

}

\maketitle
\begin{abstract}
As a distributed hash table (DHT) routing overlay, Skip Graph is used in a variety of peer-to-peer (P2P) systems including cloud storage, social networks, and search engines. The overlay connectivity of P2P systems is negatively affected by the arrivals and departures of nodes to and from the system that is known as \textit{churn}. Preserving connectivity of the overlay network (i.e., the reachability of every pair of nodes) under churn is a performance challenge in every P2P system including the Skip Graph-based ones. The existing decentralized churn stabilization solutions that are applicable on Skip Graphs have intensive communication complexities, which leave them unable to provide a strong overlay connectivity, especially under high rates of churn. 

In this paper, we propose \textit{\churnStabilization}, a fully decentralized churn stabilization mechanism for Skip Graphs that provides drastically stronger overlay connectivity without changing the asymptotic complexity of the Skip Graph in terms of storage, computation, and communication. We also propose the Sliding Window De Bruijn Graph (\textit{\swdbg}) as a tool to predict the availability of nodes with high accuracy. 
Our simulation results show that in comparison to the best existing DHT-based solutions, \textit{\churnStabilization} improves the overlay connectivity of Skip Graph under churn with the gain of about $\successRatioGain$ times. A Skip Graph that benefits from \textit{\churnStabilization} and \textit{\swdbg} is about $\searchTimeGain$ times faster on average in routing the queries under churn compared to the best existing solutions. 
We also present an adaptive extension of \textit{\churnStabilization} to be applied on other DHTs, for example Kademlia. 
\end{abstract}
\section{Introduction}
Skip Graph \cite{aspnes2007skip} is a structured overlay where nodes can efficiently perform searches for other nodes or their data objects in a fully decentralized manner. To perform the searches, each node needs to know only a few other nodes of the system, namely the node's neighbors. Using the neighboring knowledge, nodes exploit a Skip Graph overlay as a distributed routing infrastructure to initiate or route a search message. Each Skip Graph node keeps its neighbors as $\{identifier, address\}$ pairs in its lookup table. 
Neighboring relationships in Skip Graphs hence resemble the general idea of the distributed hash tables (DHTs) and yield Skip Graph being known as a DHT-based routing overlay. Because of fast searching, load balancing, and scalability, the Skip Graph is considered as a suitably structured DHT overlay for distributed services such as P2P cloud storage \cite{udoh2011cloud, batrashort, shabeera2012authenticated, hassanzadeh2016awake, hassanzadeh2015locality, hassanzadeh2016laras, hassanzadeh2018decentralized, boshrooyehguard}, and likewise, it can be applied as an alternative overlay in DHT-based applications.

Nodes in a P2P system switch between offline and online states intermittently. 
Switching to an offline state is considered as a departure from the system. 
A departed node may come back at a later time and start another online session or may leave the system permanently. Such behavior of dynamic arrivals and departures of the nodes to and from the P2P system, respectively, is referred as \textit{churn}. 
Churn jeopardizes the connectivity of the overlay network, which we define as reachability of every pair of nodes through the overlay. The compromised overlay connectivity results in search failure, inconsistent search results, and out-dated lookup table entries.

The existing churn stabilization solutions that are applicable on a Skip Graph aim to augment the overlay network by increasing the communication complexity of Skip Graph from $O(\log{\systemCapacity})$ to $O(\log^{2}{\systemCapacity})$ \cite{jacob2014skip+}, distorting the Skip Graph topology \cite{goodrich2006rainbow} that makes it inapplicable on many scenarios (e.g., locality-awareness \cite{toda2017autonomous}), tweaking the size of lookup tables based on the churn rate of the underlying system with minimum consideration of nodes' availability \cite{herrera2007modeling}, frequently probing the online status of each neighbor \cite{medrano2015performance, rhea2004handling, li2004comparing} that applies a constant communication complexity to the system, or allocating a set of backup neighbors that are contacted alternatively in the event of an unnoticed departure of a neighbor (i.e., the neighbor goes offline without informing the others) \cite{jacob2014skip+, maymounkov2002kademlia, heck2017evaluating}. The common downside of all the existing applicable churn stabilization solutions on Skip Graph overlay is that their objective function does not consider node's position in the overlay network, query latency, communication cost, and node's availability all together, and sacrifices at least one of them in favor of the rest, which negatively affects the query processing and response time of the system.

To preserve the structural integrity, as well as the routing functionality of the Skip Graph-based P2P overlays under churn, \textbf{we propose \textit{\churnStabilization}, a fully decentralized churn stabilization mechanism for the Skip Graph-based P2P overlays}. \textit{\churnStabilization} is a backup-based churn stabilization solution that utilizes backup neighbors, and provides scoring mechanisms based on their positioning in the overlay, routing latency, communication cost, as well as neighbor's availability probability. \textit{\churnStabilization} does not change the asymptotic complexity of the Skip Graph in terms of communication, computation, and storage. As an independent contribution, we also propose \underline{S}liding \underline{W}indow \underline{D}e \underline{B}ruijn \underline{G}raph (\textit{\swdbg}), a fine-grained mechanism to predict the availability probability of the nodes under churn. \textit{\churnStabilization} uses \textit{\swdbg} as a tool to predict the availability of the nodes. Compared to the existing solutions, by benefiting from \textit{\churnStabilization} and \textit{\swdbg}, a node can efficiently route the search messages in the absence of its online lookup table's neighbors with the {maximized} average success ratio as well as the {minimized} average search latency. We define the average success ratio of the searches as the ratio of successfully completed searches over all the initiated searches in the system, and define the average search latency as the time it takes for the searches to be routed to the search targets or to be declared as failures. 
Since Skip Graph can be utilized as a DHT alternative, any DHT-based application can benefit from \textit{\churnStabilization} and \textit{\swdbg}.

We consider two main goals in the design of \textit{\churnStabilization}; maintaining both the connectivity of overlay (i.e., the success ratio of searches) and overlay's speed (i.e., search latency) for a Skip Graph that undergoes churn. To increase the probability of successful searches under churn, our proposed \textit{\churnStabilization} employs backup neighbors that are contacted alternatively upon detection of an unnoticed departure of an overlay neighbor (i.e., a lookup neighbor). Using \textit{\churnStabilization}, each node keeps its backup neighbors in a memory space, named \textit{backup table}, with the same asymptotic space complexity as the lookup table (i.e.,  $O(\log{\systemCapacity})$). Although larger than $O(\log{\systemCapacity)}$ backup tables seem more successful on routing the searches by providing more alternatives, they increase the communication complexity needed for routing the search queries. The increased communication complexity increases the overall search time and applies additional communication overhead for maintenance that is not bandwidth friendly and congests the system in larger scales. As a general design strategy, \textit{\churnStabilization} gives more priority on minimizing the overall routing time under churn than maximizing the overlay connectivity. This is in contrast to the existing solutions that solely aim to maximize the overlay connectivity with the minimum attention to the routing time.  
As a supporting example, a system with the search success ratio of $0.7$ but average search latency of $10$ seconds is preferable over its counterpart with success ratio of $0.9$ but average search latency of $30$ seconds. As a failed search in the former system is highly probable to be successful at the second trial, resulting the overall average search time of $20$ seconds, which is still $1.5$ times faster in terms of query processing time than the latter. 

Compared to the existing DHT-based solutions using backup tables \cite{maymounkov2002kademlia, medrano2015performance, jacob2014skip+, li2004comparing, trifa2014effects}, \textit{\churnStabilization} offers a more delicate heuristic in terms of the search path length and search path latency that improves both the overlay connectivity and response time of the system, respectively. Additionally, \textit{\churnStabilization} operates without any maintenance communication overhead required. To improve the search latency, \textit{\churnStabilization} works on top of an availability prediction scheme that helps to consider a precedence for backup neighbors based on their availability probability, and contact the most likely online ones first. Our proposed \textit{\swdbg} executed by each node predicts the node's availability probability, and serves \textit{\churnStabilization} with this need.

The contributions of this paper are as follows:
\begin{itemize}
    \item We propose \textit{\churnStabilization}, which is a fully decentralized churn stabilization protocol for the Skip Graph-based P2P overlays. 
    
    \item As an independent contribution, we propose the Sliding Window De Bruijn Graph (\textit{\swdbg}) that provides an accurate estimate of the availability probability of the nodes.
    
    \item We provide an analytical model to predict the behavior of \textit{\churnStabilization} under uniform failure model, and to estimate the proper backup size that maximizes the performance of the system.
    
    \item We extended the Skip Graph simulator, SkipSim \cite{skipsim}, implemented, and simulated the best known decentralized churn stabilization and availability prediction mechanisms under churn, and compared with our \textit{\churnStabilization} and \textit{\swdbg}.
    
    \item Our simulation results show that compared to the best existing solutions that are applicable to the Skip Graph, \textit{\churnStabilization} improves the search success ratio of the Skip Graph overlay with the gain of about \textbf{$\successRatioGain$} times. Likewise, \textit{\swdbg} improves the availability prediction of the nodes with the gain of about \textbf{$\predictionGain$}. The search process of a Skip Graph-based P2P system that benefits from \textit{\churnStabilization} and \textit{\swdbg} is about $\searchTimeGain$ times faster on average compared to the best existing solutions.\\
\end{itemize}

In Section \ref{section:interlace_preliminaries} we describe the structure of Skip Graph, its typical search for numerical ID protocol, and preliminaries such as De Bruijn Graph and churn model. In Section \ref{section:interlace_model} we state our system model. Our proposed \textit{\swdbg} and \textit{\churnStabilization} are presented in Sections \ref{section:interlace_sliding_bruijn_graph} and \ref{section:interlace_interlace}, respectively. The related works are surveyed in Section \ref{section:interlace_relatedworks}. Our simulation setup followed by analytical and performance results are presented in Sections \ref{section:interlace_simulation} and \ref{section:interlace_results}. We conclude the paper in Section \ref{section:interlace_conclusion}.

\section{Preliminaries}
\label{section:interlace_preliminaries}
\subsection{Skip Graph}
\begin{figure}
\centering
\includegraphics[width=\linewidth]{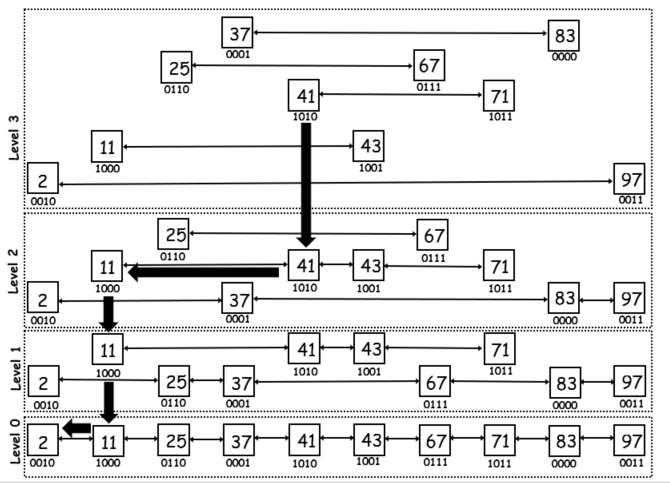}
\caption{An example Skip Graph with $10$ nodes and $4$ levels. Nodes are visualized by squares with numerical IDs enclosed and name IDs beneath. Example search for numerical ID of $2$ that is initiated by node $41$ is depicted by the thick arrows.} 
\label{figure:interlace_skipgraph}
\end{figure}

\textbf{Structure:} An example Skip Graph \cite{aspnes2007skip} with $10$ nodes and $4$ levels is represented by Figure \ref{figure:interlace_skipgraph}. In general, a Skip Graph with $\systemCapacity$ nodes has $O(\log{\systemCapacity})$ levels that are numbered starting from $0$ in a bottom-up manner. Each Skip Graph node has exactly one element in each level, and is identified with a name ID and a numerical ID. Name IDs are binary strings of size $O(\log{\systemCapacity})$ bits, and numerical IDs are non-negative integers. In Figure \ref{figure:interlace_skipgraph}, elements of a node are represented by squares, with numerical IDs enclosed and name IDs are located beneath each element. 

Level $0$ of a Skip Graph has exactly one distributed list with nodes that are sorted in ascending order. Distributed list means that there is no central entity (e.g., server) that is supposed to keep the list partially or as the whole; instead, each list's element keeps the address of its successor and predecessor. In level $i > 0$, there exists $2^{i}$ lists, where nodes in each list have a common prefix of at least $i$ bits long in their name IDs. For example, in Figure \ref{figure:interlace_skipgraph}
name IDs of $\underline{0}010$, $\underline{0}110$, $\underline{0}001$, $\underline{0}111$, $\underline{0}000$, and $\underline{0}011$, all coexist in the same list at level $1$ of the Skip Graph since all their name IDs start with $0$ prefix. Likewise, name IDs of
$\underline{00}10$, $\underline{00}01$, $\underline{00}00$, and $\underline{00}11$ are located in the same list at level $2$ due to their $2$-bit common prefix of $00$. 
However, there exists no name ID in the Skip Graph with the prefix of $11$, which makes the corresponding list with the prefix of $11$ on level $2$ empty. 
Without loss of generality, in this study, we assume the uniqueness of name IDs and numerical IDs and hence we identify a node with either its name ID or numerical ID e.g., by node $43$ we mean the node that holds the numerical ID of $43$ and the name ID of $1001$.

\begin{figure}
\centering
\includegraphics[scale= 0.4]{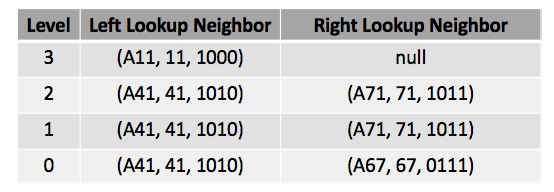}
\caption{Lookup table of node $43$ from Figure \ref{figure:interlace_skipgraph}.} 
\label{figure:interlace_lookup}
\end{figure}

In a Skip Graph-based P2P overlay, each peer from the real world is represented by a Skip Graph node. 
The numerical ID of each node is the hash value of its corresponding peer's IP address. We assume the name IDs of nodes are generated by a locality-aware name ID strategy e.g., LANS \cite{hassanzadeh2018decentralized}. With locality-aware name IDs, the latency between two nodes in the underlying network is an inverse function of the length of their name IDs' common prefix in the Skip Graph overlay, i.e., longer common prefix conveys lower latency.
In a Skip Graph-based P2P overlay, a node is supposed to only know its directly connected predecessor and successor at each level, which are called its left and right neighbors on that level, respectively.
A Skip Graph node keeps its neighboring information locally as $($\textit{address, numerical ID, name ID}$)$ tuples in a table, which is called the lookup table of that node as its local view of the system. The lookup table of node $43$ from Figure \ref{figure:interlace_skipgraph} is illustrated in Figure \ref{figure:interlace_lookup} where $Axx$ is the (IP) address of the node with numerical ID of $xx$.

\textbf{Search for numerical ID: }
Search for numerical ID is a fully decentralized search protocol of a Skip Graph, where a node named the \textit{search initiator} starts the search for a \textit{target numerical ID}. Considering a Skip Graph-based P2P system as a distributed database of $(address, numerical\_ID, name\_ID)$ records, the search for numerical ID is analogous to a distributed $get()$ that retrieves the address of the node with the closest numerical ID to the target numerical ID. As a convention, the search for numerical ID always returns the address of the node with the greatest numerical ID that is less than or equal to the target numerical ID. In the exceptional case where the target numerical ID is less than all the numerical IDs of the Skip Graph, the address of the lowest numerical ID that is greater than the target numerical ID is returned. 
Nodes route a search message based on their lookup table information. Although nodes may leverage their cached information from previous routings to expedite the ongoing routing tasks, nevertheless, the existence of lookup table guarantees an asymptotic communication complexity of $O(\log{\systemCapacity})$ in expectation with high probability for processing a single search for a numerical ID in a fully decentralized manner \cite{aspnes2007skip}. It is worth mentioning that the lookup table of each node is as big as the number of Skip Graph's levels i.e., $O(\log{\systemCapacity})$.

The thick arrows of Figure \ref{figure:interlace_skipgraph} show an example search for numerical ID with node $41$ as the search initiator and $2$ as the target numerical ID. A search for numerical ID always starts from the top-most level of the search initiator. In this example, since the target numerical ID of $2$ is less than the search initiator's numerical ID (i.e., $41$), the search messages are always routed towards the left. The horizontal thick arrows correspond to the routed search messages to the left neighbors, while the vertical ones represent the internal computation of nodes stepping down in their lookup table. On a certain level, the search message is forwarded repeatedly leftward while the left neighbor's numerical ID is greater than or equal the search target. If there is no such neighbor on the left, the node jumps down one level in its lookup table and checks the eligibility of its lower level's neighbor in the search direction. On receiving the search message at level $0$, node $2$ realizes that it holds the exact target numerical ID as the search message and hence introduces itself as the search result to the search initiator node $41$.

\subsection{De Bruijn Graph (DBG):}
De Bruijn Graph \cite{de1946combinatorial} (DBG) is a pattern recognition data structure that is also used as a tool to detect and extract the availability pattern of a single node \cite{mickens2006exploiting, mickens2006improving}. In this paper we utilize a slightly modified version of DBG as follows. We identify a DBG by a \textit{state size} of $\bruijnStateSize$ bits and $2^\bruijnStateSize$ vertices that are labeled by the $\bruijnStateSize$-bit binary representation of all integers in $[0, 2^{\bruijnStateSize}-1]$ range. A vertex with the binary label of $b_{1}b_{2}...b_{\bruijnStateSize}$ has exactly two outgoing edges to the vertices associated with $\underline{b_{2}}b_{3}...\underline{0}$ and $\underline{b_{2}}b_{3}...\underline{1}$. In representation of the availability behaviour of a node with a DBG, the time is divided into time slots with fixed identical size. Each DBG vertex represents the availability history of the node within the last $\bruijnStateSize$ time slots, 
with $1$ corresponding to an online state, and $0$ corresponding to an offline state. The rightmost and leftmost bits of each state represent the newest and oldest availability status of the node within a window of $\bruijnStateSize$ time slots, respectively. Having the availability history of a node within the last $\bruijnStateSize$ time slots as $b_{1}b_{2}...b_{\bruijnStateSize}$, the outgoing edges to $b_{2}b_{3}...\underline{0}$ and $b_{2}b_{3}...\underline{1}$ denote the availability status of node in the $\bruijnStateSize+1^{st}$ slot, which is represented by the rightmost bit (i.e., the underlined ones). The outgoing edge $b_{1}b_{2}...b_{\bruijnStateSize} \rightarrow b_{2}b_{3}...1$ is associated with $p_{b_{1}b_{2}...b_{\bruijnStateSize}}^{1}$ i.e., the probability of being online in the $\bruijnStateSize+1^{th}$ time slot given the history of $b_{1}b_{2}...b_{\bruijnStateSize}$ in the last $\bruijnStateSize$ time slots. Similarly, $p_{b_{1}b_{2}...b_{\bruijnStateSize}}^{0}$ relates to the edge $b_{1}b_{2}...b_{\bruijnStateSize} \rightarrow b_{2}b_{3}...0$, and represents the probability of being offline in the $\bruijnStateSize+1^{th}$ time slot given the availability history of $b_{1}b_{2}...b_{\bruijnStateSize}$ for the node in the last $x$ time slots. The probabilities are taken over all the time slots a specific state is visited, which yields $p_{state}^{0} + p_{state}^{1} = 1$ for every $state$ of a DBG.

\subsection{Churn Model}
P2P overlays are dynamic with respect to the time i.e., nodes frequently switch between online and offline states. 
Such dynamic aspect of the system is described by a churn model \cite{stutzbach2006understanding}. A churn model is identified with a session length and an inter-arrival time distribution. Session length distribution characterizes the online duration of the nodes in the system. The inter-arrival time distribution characterizes time between the start of two consecutive online sessions of nodes in the system. 
\section{System Model and Scenario}
\label{section:interlace_model}
\textbf{Model:} We consider the Skip Graph as an application layer protocol that is executed independently by the peers in an honest manner i.e., each peer follows the exact protocol without deviation and represents a Skip Graph node. The independent executions of Skip Graph protocol by the participating peers shapes a P2P overlay that is constructed by joining the first peer to the system and grows over the time by joining other peers. In particular, peers use the insertion algorithm of the Skip Graph to join the system as Skip Graph nodes \cite{aspnes2007skip}. After joining the system, nodes frequently perform the search for numerical IDs to find each other and resources, help other nodes joining the system, or to perform other P2P tasks e.g., replication \cite{hassanzadeh2016awake, ramachandran2012decentralized, hassanzadeh2016laras}.
We define the \textit{system capacity} $\systemCapacity$ as the number of registered users to the Skip Graph-based P2P system which is constant despite the churn of nodes. In this paper, we consider the system capacity as the smallest power of two that is greater than or equal to the total number of registered nodes in the system. We define the \textit{timeout failure} as the situation where a node does not hear from its offline neighbor within a certain time interval after routing a search message to it. We consider the time interval duration that can trigger a timeout failure as the function of the round trip time between the node and its neighbor.

\textbf{Scenario:}
Using our proposed \textit{\swdbg} (see Section \ref{section:interlace_sliding_bruijn_graph}), 
each online node computes its own availability probability at the end of each time slot and piggybacks it alongside its address, name ID, and numerical ID on all the search messages it routes or initiates. On receiving a search message to route, the node updates its backup table with the piggybacked availability information of other nodes. To do such an update, upon a message reception, the node invokes the \textit{\backupUpdate} event handler of our proposed \textit{\churnStabilization}. As described in Section \ref{section:interlace_preliminaries}, a node routes a search message by forwarding it to one of the lookup neighbors that is placed in the level and direction of the search. However, due to the churn, the selected lookup neighbor may be offline, and unable to receive the search message. We presume an associated timer for every forwarded message with the expiration time as a function of the round trip time between the node and its corresponding neighbor. If no (TCP) acknowledgment is received from the neighbor before the timer expires, that neighbor is considered as offline, and a timeout failure happens. On timeout failures, the node invokes the \textit{\backupResolve} event handler of \textit{\churnStabilization}, which returns back an online candidate node from the backup table that is consistent with the search path, and the search message is redirected to this online candidate.

\section{Sliding Window De Bruijn Graph (\textit{\swdbg})}
\label{section:interlace_sliding_bruijn_graph}
\subsection{\textbf{Overview}}
\begin{figure}
\centering
\includegraphics[width=\linewidth]{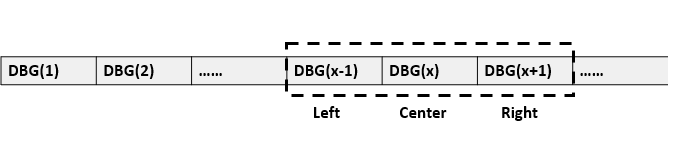}
\caption{The general form of Sliding Window De Bruijn Graphs (\textit{\swdbg}) with the current state window adjusted on DBG($x-1$), DBG($x$), and DBG($x+1$), which are denoted by Left, Center, and Right DBGs, respectively.} 
\label{figure:interlace_swdbg}
\end{figure}

Choosing a proper state size for DBG predictors is a challenge as large state sizes enforce a noise over the DBG that increases the prediction error \cite{mickens2006exploiting}, while small state sizes fail DBG to completely capture the availability behavior of the nodes. Furthermore, DBG has exponential asymptotic space and time complexities as a function of its state size. Therefore, finding the smallest state size that provides an acceptable level of prediction accuracy lets the nodes to operate efficiently in both space and time.  
In order to find the proper state size of DBGs and predict the availability probability of a node in an adaptive manner, we propose the Sliding Window De Bruijn Graph (\textit{\swdbg}). By adaptive we mean that in contrast to the existing DBG-based predictors \cite{mickens2006exploiting, mickens2006improving}, which have a fixed state size, \textit{\swdbg} continuously moves towards the state size that describes the node's instantaneous availability status the best. 
As shown by Figure \ref{figure:interlace_swdbg}, \textit{\swdbg} is a list of DBGs that are represented by DBG($\bruijnStateSize$) where $\bruijnStateSize$ is the state size of the corresponding DBG. The list is started by DBG($1$), and the state size of DBGs increases by one moving from left to the right. Despite this long list of DBGs, however, an \textit{\swdbg} is required to only keep three consecutive DBG instances on its memory space, and operate on those accordingly. This set of three consecutive DBGs is called as the \textbf{current state window} that is represented by a dashed rectangle in Figure \ref{figure:interlace_swdbg}.
Each node updates the current state window's DBGs of its \textit{\swdbg} once within each time slot, and computes and piggybacks its stationary online probability on all the search messages it routes or initiates. We define the \textbf{stationary online probability} of a node as the probability of the node being online after infinitely many time slots elapsed (i.e., at the time $t \rightarrow \infty$). 
The stationary online probability of a node is updated frequently (i.e., once every time slot) to reflect the possible changes in the availability behavior. After each update, \textit{\swdbg} may move the current state window to the left or right if it realizes that a smaller or bigger state size may describe the availability status of the node with lower prediction error. We assume that an offline node updates its \textit{\swdbg} for all its offline time slots right at the end of its first new online time slot at the system upon arrival.

\subsection{\textbf{Stationary Online Probability}}
We model each DBG($\bruijnStateSize$) of a \textit{\swdbg} with a single Markov Chain process \cite{bertsekas2002introduction}, by representing each vertex $i$ of DBG (i.e., $i = b_{1}b_{2}...b_{\bruijnStateSize}$) with a Markov Chain state where $\bruijnStateSize$ is the state size of DBG. The transition matrix of each DBG is represented by $P$, and shown by Equation \ref{eq:interlace_transition}. $p_{i,j}$ is the transition probability from state $i$ to state $j$, and is determined by Equation \ref{eq:interlace_markov}. 
$\stateSet$ denotes the set of DBG($\bruijnStateSize$)'s states, and constitute of all $\bruijnStateSize$ bits binary strings. 
As presented in Section \ref{section:interlace_preliminaries}, for state $i = b_{1}b_{2}...b_{\bruijnStateSize}$, $p_{i}^{0}$ and $p_{i}^{1}$ denote the transition probabilities of $b_{1}b_{2}...b_{\bruijnStateSize} \rightarrow b_{2}b_{3}...0$ and $b_{1}b_{2}...b_{\bruijnStateSize} \rightarrow b_{2}b_{3}...1$, respectively. 

\begin{equation}
    P = \{p_{i,j}| 1 \leq i,j \leq \bruijnStateSize\}
    \label{eq:interlace_transition}
\end{equation}

\begin{equation}
    p_{i,j}= 
\begin{dcases}
    p_{i}^{0} ,& \text{if } j = b_{2}...b_{\bruijnStateSize}0\\
    p_{i}^{1} ,& \text{if } j = b_{2}...b_{\bruijnStateSize}1\\
    0,              & \text{Otherwise}
\end{dcases}
\label{eq:interlace_markov}
\end{equation}

\noindent When the Markov Chain analogy of a DBG($\bruijnStateSize$) is Ergodic (i.e., each state is accessible from every other state), the system that is illustrated by Equation \ref{eq:interlace_stationary} comes to a unique answer in the form of $\{\pi_1, \pi_2, ..., \pi_\bruijnStateSize\}$. $\pi_{i}$ denotes the stationary probability of state $i$, and represents the probability of visiting state $i$ after infinitely many state transitions independent of the initial state.  

\begin{equation}
    \begin{dcases}
    1 &= \sum_{i \in \stateSet} \pi_{i}\\
    \pi_{i} &= \sum_{j \in \stateSet} \pi_{j} \times p_{j,i}, \forall j \in \stateSet\\
    \end{dcases}
    \label{eq:interlace_stationary}
\end{equation}

\noindent We represent the stationary online probability of an Ergodic DBG($\bruijnStateSize$) ($\bruijnStateSize \geq 1$) by $sop_{\bruijnStateSize}$. As shown by Equation \ref{eq:interlace_sop}, $sop_{\bruijnStateSize}$ corresponds to the aggregated stationary probabilities of all DBG's states $i$ that end with $1$. Since the rightmost bit of each state corresponds to the most recent availability status,
the states that end with $1$ represent an online status of the node in its availability history. Summing up the stationary probabilities of all those states results in the stationary online probability of the node. 

\begin{equation}
    sop_{\bruijnStateSize} = \sum_{\{i \in \stateSet | i = b_{1}b_{2}...1\}} \pi_{i}
    \label{eq:interlace_sop}
\end{equation}

\subsection{\textbf{Algorithm Description}}
As shown by Figure \ref{figure:interlace_swdbg}, at any point in time, the current state window has three DBGs that are denoted by DBG($\bruijnStateSize-1$), DBG($\bruijnStateSize$), and DBG($\bruijnStateSize+1$), $\bruijnStateSize \geq 2$, and called the Left, Center, and Right DBGs, respectively. A node initializes its instance of \textit{\swdbg} by adjusting the Left DBG of the current state window on DBG($1$). While the node is online, it updates the current state window with its own availability status, by invoking $stateUpdate$ algorithm at the end of each time slot. As shown by Algorithm \ref{intelace:alg_onlinestateUpdate}, the inputs to $stateUpdate$ are the current state window $cw$ (i.e., the collection of $3$ DBGs; Left, Center, and Right), as well as the $status$ of the node, which is $1$ if the node is online, and $0$ otherwise. On receiving the inputs, $stateUpdate$ updates the DBGs inside the current state window by a call to their $update$ routine that updates the current state of DBG with the $status$ bit, updates the proper state transitions' probabilities, and returns the stationary online probability upon existence. If the modeled Markov Chain of a DBG does not show \textit{ergodicity} \cite{bertsekas2002introduction}, $update$ returns either $0$ or $1$ depending on the existence of an offline or online absorbing state, respectively (Algorithm \ref{intelace:alg_onlinestateUpdate}, Lines \ref{intelace::alg_onlinestateUpdate:beginning}-\ref{intelace::alg_onlinestateUpdate:updateDone}). In other words, in our system scenario, non-ergodicity happens when there exists an absorbing online or offline state, and hence there exists no stationary distribution for the modeled Markov Chain. By an absorbing offline or online state, we mean an absorbing state that ends with $0$ or $1$, respectively.

$updateStatus$ computes the prediction error of the sliding window's DBGs that is determined as the absolute value of the difference between their $sop$ probability and their availability fraction within a state size window of last time slots. For example, if one DBG($3$) predicts the stationary online probability as $0.2$ while the availability status of the node in the window of last $3$ time slots is $101$, the prediction error is computed as $|0.2 - \frac{2}{3}| = 0.44$, which is a noticeable prediction error. The strictly decreasing prediction errors of current state window's DBGs towards either the right (i.e., $predErr_{Left} > predErr_{Center} > predErr_{Right}$), or the left (i.e., $predErr_{Left} < predErr_{Center} < predErr_{Right}$) implies a prediction accuracy improvement on moving the current state window towards the right or left direction, respectively. A right movement is done by dropping the $Left$ DBG out of the current state window and replacing its pointer with $Center$ DBG. The $Center$ DBG is also replaced by the $Right$ DBG, and finally the $Right$ DBG's state size is increased by one bit via performing a call on its $enlarge$ function that does the enlargement by mapping each state $b_{1}b_{2}...b_{\bruijnStateSize}$ to the two states $b_{1}b_{2}...b_{\bruijnStateSize}0$ and $b_{1}b_{2}...b_{\bruijnStateSize}1$ while preserving the associated probabilities of the states. A left movement is done similar to the right movement except that the function $shrink$ of the $Left$ DBG is called that maps each two states  $b_{1}b_{2}...b_{\bruijnStateSize-1}0$ and $b_{1}b_{2}...b_{\bruijnStateSize-1}1$ to a single  $b_{1}b_{2}...b_{\bruijnStateSize-1}$ state in the new shrunk $Left$ DBG. We compute the probability of each shrunk state as the average probability of the states that are mapped to it (Algorithm \ref{intelace:alg_onlinestateUpdate}, Lines \ref{intelace::alg_onlinestateUpdate:errorComputation}-\ref{intelace::alg_onlinestateUpdate:movementsDone}).  
The $updateStatus$ function terminates its task by returning back the stationary online probability of the current state window that is defined as the stationary online probability of its DBG with minimum prediction error (Algorithm \ref{intelace:alg_onlinestateUpdate}, Line \ref{intelace::alg_onlinestateUpdate:endline}). With this approach of shrinking and enlarging the current state window's DBGs, our proposed \textit{\swdbg} adaptively chooses the DBG size that best describes the availability behavior of the node, which is in contrast to the conventional strategy of traditional DBG on having a fixed-size state size independent of the node's availability behavior.

\begin{algorithm}
\SetAlgoLined
\KwIn{DBG[] $cw$, bit $status$}
     \tcp{updating the current state window with input status}  
     \For{$i \in \{Left, Center, Right\}$}
     {
           \label{intelace::alg_onlinestateUpdate:beginning}
        $sop_{i} = cw[i].update(status)$\; 
     }
    \label{intelace::alg_onlinestateUpdate:updateDone} 

      \tcp{computing the prediction errors}  

    \For{$i \in \{Left, Center, Right\}$}
    {
          \label{intelace::alg_onlinestateUpdate:errorComputation} 
     $predErr_{i} = \left|status - sop_{i}\right|$\;
    }

      \tcp{checking for the enlarge or shrink}        
      \While{$predErr_{Left} > predErr_{Center} >  predErr_{Right}$}
      {
         \label{intelace::alg_onlinestateUpdate:movementsStart} 
        Slide the current window to the right\;
        $cw[Right] = cw[Right].enlarge()$\;
      }
      \While{$predErr_{Left} < predErr_{Center} < predErr_{Right}$}
      {
       Slide the current window to the left\;
        $cw[Left] = cw[Left].shrink()$\;
      }
     
    \label{intelace::alg_onlinestateUpdate:movementsDone} 
     
       return $cw[\argmin_{predErr} \{cw\}].sop$\;
       \label{intelace::alg_onlinestateUpdate:endline} 
 
\caption{stateUpdate}
\label{intelace:alg_onlinestateUpdate}
\end{algorithm}

\section{\textit{\churnStabilization}}
\label{section:interlace_interlace}
\subsection{\textbf{Overview}}
\textit{\churnStabilization} is a fully decentralized application layer churn stabilization protocol that is executed independently by every node of a Skip Graph-based P2P overlay. A node executes \textit{\churnStabilization} upon facing a timeout failure (see Section \ref{section:interlace_model}) on routing a search message based on its lookup table.
We denote the node that executes an instance of \textit{\churnStabilization} by \textit{executor}.
In other words, any node running \textit{\churnStabilization} is named as an executor, and this does not imply the restriction to any special node like a super node.
The main goal of \textit{\churnStabilization} is to recover a search query from a timeout failure on its current path. \textit{\churnStabilization} assumes an additional $O(\log{n})$ storage of the neighboring information for the executor that is called \textit{backup table}, which is a common trait among the existing solutions \cite{maymounkov2002kademlia, medrano2015performance, li2004comparing, trifa2014effects}.
The executor holds the collected availability information of some other nodes of Skip Graph on its backup table in a level-wise manner similar to the lookup table. On referring to the backup table,  \textit{\churnStabilization} finds the best routing candidate and redirects the search message to it. \textit{\churnStabilization} chooses the best routing candidate based on the numerical ID distance to the search target (i.e., number of intermediate nodes on the path), the locality-aware name ID similarity with the executor (i.e., common prefix) that corresponds to a expected latency in the underlying network, and the stationary online probability of the candidate. 
On redirecting the search message to the best candidate, if no acknowledgement is received within a certain time interval, the best candidate is presumed offline, and \textit{\churnStabilization} moves to the next best routing candidate. 

\subsection{\textbf{Backup table}}
The backup table resembles the lookup table in structure, except that instead of storing a single node information at every entry (i.e., cell), each entry of a backup table represents a set of nodes' information. Hence, each level of a backup table constitutes of two \textbf{entry sets} (i.e., one at the left and one at the right), each representing a set of nodes' information. 

\begin{figure*}
\centering
\begin{minipage}[c]{.48\textwidth}
\centering
     \includegraphics[scale= 0.3]{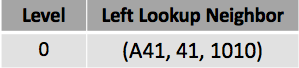}
    \caption*{(a) Left lookup table entry at level $0$}
\end{minipage}

\begin{minipage}[c]{.48\textwidth}
\centering
    \includegraphics[scale= 0.3]{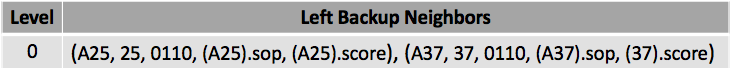}
    \caption*{(b) Left backup table entry set at level $0$}
\end{minipage}
\caption{A comparison between a lookup table entry of node $43$ in the sample Skip Graph of Figure \ref{figure:interlace_skipgraph}, and its corresponding potential backup table entry set at the same level and direction.}
\label{interlace:fig_lookup_backup}
\end{figure*}

\textit{\churnStabilization} changes the size of each set adaptively with the overall number of backup neighbors being within the range of $[0,\backupSize]$ where $\backupSize$ is a system-wide constant named as the \textit{backup size}. This implies that backup table has the same memory complexity of $O(\log{\systemCapacity})$  as the lookup table.
Each entry holds the information of a single node in the form of $(address, numID, nameID, sop, score)$ tuple where $address$ represents the (IP) address of the node in the underlying network. $numID$ and $nameID$ represent the numerical and name IDs of the node, respectively. The stationary online probability of the node (see Section \ref{section:interlace_sliding_bruijn_graph}) is denoted by $sop$, and $score$ is a real number that is frequently updated and used by \textit{\churnStabilization} to select the best routing candidate. Figure \ref{interlace:fig_lookup_backup} shows the left lookup table entry of node $43$ at level $0$ from the sample Skip Graph of Figure \ref{figure:interlace_skipgraph}, and a potential corresponding backup table entry set to it at the same level and direction. Note that although backup table entry set of Figure \ref{interlace:fig_lookup_backup} holds two neighbors, nevertheless, these two neighbors are solely for the sake of clarification, and in practice a backup table entry set is able to hold an arbitrary number of neighbors limited by the overall size $\backupSize$ of the backup table.

\subsection{\textbf{Algorithm Description}}
\textit{\churnStabilization} consists of two event handlers: \textit{\backupUpdate} and \textit{\backupResolve} that are called by the executor on the events of receiving a search message to route, and timeout failure, respectively. 

\subsubsection{\textbf{Updating backup table (\textit{\backupUpdate}):}} 
\textbf{Identifying the proper entry set:}
\noindent In our system model, we assume that while a node is online it computes its stationary online probability by applying \textit{\swdbg} on its availability history, and piggybacks its (IP) address, numerical ID, name ID, and stationary online probability on the search messages it routes or initiates (as described in Section \ref{section:interlace_sliding_bruijn_graph}). On reception of a search message that contains the piggybacked information of other nodes, the executor invokes the \textit{\backupUpdate} event handler of \textit{\churnStabilization} on the set of piggybacked information.
On receiving the inputs, \textit{\backupUpdate} inserts each piggybacked element $e$ of the message that does not correspond to a lookup table neighbor, into a proper entry set of the executor's backup table. The proper set is identified by the direction and level. The direction is determined based on the numerical ID position of the executor with respect to the piggybacked element's numerical ID i.e., right direction if $e.numID > executor.numID$, and left direction otherwise. The level corresponds to the common prefix length between the executor's and element $e$'s name IDs. If an older version of $e$ (i.e., a version with possibly different stationary online probability) resides in the backup table, \textit{\backupUpdate} overwrites it with the new element. By an older version, we mean an element $e$ that exists in the backup table due to the \textit{\backupUpdate} invocation on previous messages that contained $e$. Otherwise, it adds the piggybacked element into the proper backup table's entry set that is identified by the level and direction.

\textbf{Scoring-based element replacement:} Before inserting a new piggybacked element into the proper backup table entry set, \textit{\backupUpdate} evaluates the size of the backup table against the maximum permissible size $\backupSize$. 
In contrast to the existing solutions that discard the oldest entry in the case that backup table is full, \textit{\churnStabilization} provides a scoring mechanism to distinguish the backup table neighbors and replaces the element with the minimum score by the new piggybacked element. The scoring-based element replacement mechanism of  \textit{\backupUpdate} is enabled upon a backup table size violation, sorts all the entries of backup table based on their updated score value as represented by Equation \ref{eq:interlace_update_scoring}, and drops the element with the minimum score value out of the backup table. This squeezes the size of backup table down to $\backupSize - 1$ element, which can accommodate the new piggybacked element.

\begin{equation}
    e.score = \frac{e.sop \times commonPrefixLength }{\left|e.numID - numID\right|}
    \label{eq:interlace_update_scoring}
\end{equation}

In Equation \ref{eq:interlace_update_scoring}, $e.score$ denotes the score that \textit{\backupUpdate} assigns to element $e$, which has a direct relation with its stationary online probability (i.e., $e.sop$) as well as the length of common prefix between the name IDs of executor and element $e$ (i.e., $commonPrefixLength$). We utilize a locality-aware Skip Graph \cite{hassanzadeh2015locality} where name IDs' common prefix length reversely reflects the underlying latency between two nodes i.e., a longer common prefix length reflects a lower latency. Hence, the score of an element is a direct function of its availability probability that is projected by its latency towards the executor node. Moreover, the score of an element is an inverse function of its numerical ID distance with respect to the executor. Following the search for numerical ID protocol (see Section \ref{section:interlace_preliminaries}), the numerical ID distance to the search target keeps decreasing as the search message proceeds towards the target. This hints that as the search path lengthens, the probability that an executor node on the path receives a search message for a target in its own numerical ID vicinity increases. Consequently, for the routing to be successfully conducted under churn, \textit{\backupUpdate} aims to keep the backup table elements in the numerical ID proximity of the executor node by assigning a higher score to the backup neighbors in shorter numerical ID distances than the others. This increases the chance of having routing candidates inline with the search direction and in the proximity of the executor's numerical ID. Once the size reduction is done, \textit{\backupUpdate} function adds the new element to the identified set by the level and direction, and returns back the updated backup table once all elements in the input search message were processed. 

\subsubsection{\textbf{Using backup table (\textit{\backupResolve}):}}
As the timeout failure event handler on routing a search message, the executor invokes \textit{\backupResolve} algorithm that is shown by Algorithm \ref{intelace:alg_resolvingFailure}. Inputs to \textit{\backupResolve} are the backup table and name ID of the executor (i.e., $backup$ and $nameID$, respectively), the search target numerical ID (i.e., $target$), the current ongoing level and direction of the search (i.e., $level$ and $dir$, respectively), and the search message itself (i.e., $msg$). The direction $dir$ is either LEFT or RIGHT. As the output, \textit{\backupResolve} returns an online routing candidate from the backup table if such online candidate exists. Otherwise, it returns $NULL$. If an online candidate is returned by \textit{\backupResolve}, the search message is redirected to that candidate by the executor.   

On receiving the inputs, \textit{\backupResolve} iterates over the $backup[level][dir]$ set, and evaluates the eligibility of each element as a routing candidate. This is done by invoking $candCheck(e, target, dir, msg)$ that evaluates each element from two aspects; being on the search direction, and not yet being visited by this search message (i.e., $e \notin msg$). In checking the consistency with the search direction, $candCheck$ returns $false$ if the direction is RIGHT (or LEFT) but the $e.numID$ is greater (or less) than the $target$. Moreover, to avoid looping on the search path, $candCheck$ returns $false$ if there exists a piggybacked node information on $msg$ that corresponds to element $e$ (i.e., $e \in msg$). If none of these violations happen, $candCheck$ returns $true$. All elements with the $candCheck$ value of $true$ are evaluated as tentative routing candidates and added to the $candidatesList$, which denotes the list of candidates. Before adding each candidate $e$ to the list of candidates, \textit{\backupResolve} scores it similar to the scoring strategy of  \textit{\backupUpdate} (Equation \ref{eq:interlace_update_scoring}), except that the numerical ID distance is evaluated with respect to the $target$, and not the executor node. In other words, to compute the score of each backup table element $e$, \textit{\backupUpdate} follows Equation \ref{eq:interlace_update_scoring} by replacing $e.sop$ with the stationary online probability of element $e$, $commonPrefixLength$ with the common prefix length between the name IDs of the executor and element $e$, and the denominator with the numerical ID distance of element $e$ to the search target. This scoring is to discriminate the candidates based on their stationary online probability, search path length to the $target$, and latency between them and executor. By the search path length, we mean the number of nodes on the path between the candidate and $target$. A lower numerical ID distance to the search target implies the lower number of intermediate nodes on the search path (i.e., a shorter search path) in expectation with high probability, which increases the success probability of search under churn as the search is more likely to hit the $target$ in the immediate subsequent steps. Likewise, under the locality-awareness assumption of the overlay, routing to a more similar name ID with respect to the executor, decreases the expected routing latency, which compensates the delay caused by timeout failure, partially. A higher score corresponds to a better candidate in terms of availability probability, closeness to the $target$ in the overlay network, and lower latency to the executor on receiving the redirected search message in the underlying network (Algorithm \ref{intelace:alg_resolvingFailure}, Lines \ref{intelace:alg_resolvingFailure:beggining}-\ref{intelace:alg_resolvingFailure:listCreated}).

After creating the list of candidates, \textit{\backupResolve} selects the candidate with the maximum score as the best routing candidate, checks its online status (e.g., by pinging it), and returns its address to the executor in the case that it is online. If the best candidate is offline, \textit{\backupResolve} removes it from both the backup table and candidates list and moves to the next best candidate. Removing offline candidates from the backup table is to address the permanent departure of nodes that have resided long enough in the system to show a good stationary online probability, but are no longer available after their departure. $candidatesList$ running out of candidates implies that there is no other online alternative to redirect the search message. This makes \textit{\backupResolve} to return $NULL$, which hints the termination of the search at that level to the executor (Algorithm \ref{intelace:alg_resolvingFailure}, Lines \ref{intelace:alg_resolvingFailure:contactingCandidatesStarted}-\ref{intelace:alg_resolvingFailure:end}). Termination at non-zero levels makes the executor to continue the search at lower levels (see Section \ref{section:interlace_preliminaries}). However, termination at level zero corresponds to the termination of the whole search, and executor returns back its own address as the search result to the search initiator. Moreover, following the \textit{\churnStabilization}'s preference on fast response under churn than full connectivity, it only refers the part of the backup table that is consistent with the search.

\begin{algorithm}
\SetAlgoLined
\KwIn{backup table $backup$, executor name ID $nameID$, search target numerical ID $target$, level of search $level$, direction of search $dir$, Message $msg$}
     \tcp{initializing the list of candidates to contact}  
     $candidatesList = \O$\;
     \label{intelace:alg_resolvingFailure:beggining} 
     \For{element $e \in backup[level][dir]$ }
     {
      \If{$e.numID == target$}
      {
        \tcp{the search target has been found}
        return $e.address$\;
      }
      \If{$candCheck(e, target, dir, msg)$}
      {
        \label{intelace:alg_resolvingFailure:candCheck}
        \tcp{assigning the score to the candidate}
        $e.score = \dfrac{e.sop \times commonPrefixLength }{\left|e.numID - target\right|}$\;
        \tcp{adding the candidate to the list}
        $candidatesList.add(e$)\;
      }
     }
     \label{intelace:alg_resolvingFailure:listCreated} 
     
     \While{$!candidatesList.isEmpty()$ }
     {
         \tcp{picking the best candidate from list}
         \label{intelace:alg_resolvingFailure:contactingCandidatesStarted} 
        $bestCandidate = \argmax_{score} \{candidatesList\}$\;
        \tcp{redirecting the search message to the best routing candidate}
         $send(msg, bestCandidate.address$\;
        \eIf{$bestCandidate.isonline()$}
        {
            break\;
        }
        {
           \tcp{removing the offline best candidate from the candidates list and backup table}
            $candidatesList.remove(bestCandidate$)\;
            $backup.remove(bestCandidate$)\; 
        }

     \label{intelace:alg_resolvingFailure:contactingCandidatesDone} 
 }

 {return NULL\;}
 \label{intelace:alg_resolvingFailure:end} 

\caption{\backupResolve}
\label{intelace:alg_resolvingFailure}
\end{algorithm}

\subsection{\textbf{Applying on other DHTs}}
\textit{\churnStabilization} is capable of being efficiently applied on other DHTs especially the prefix-based ones like Kademlia \cite{maymounkov2002kademlia}, 
which resembles Skip Graph in prefix-based binary string identifiers. To apply \textit{\churnStabilization} on such DHTs, each node needs to allocate a backup table and
apply some minor modifications to the event handlers of \textit{\churnStabilization} i.e., \textit{\backupUpdate} and \textit{\backupResolve}. For \textit{\backupUpdate}, determining the level and direction of a new element to be inserted in the backup table needs to be changed based on the architecture of the DHT. For example, in Kademlia where the search messages are routed based on the XOR distance between the executor's and the target's identifiers, identifying the proper backup set is done by the XOR distances, and without the need to determine any direction. As each backup set is identified by an identifier prefix, the proper backup set for a new element is the one with the minimum XOR distance to its identifier. Also, the scoring-based element replacement part of \textit{\backupUpdate} is modified as shown by Equation \ref{eq:interlace_update_scoring_kademlia} where $e.id$ and $id$ are the identifiers of the piggybacked element $e$ and executor, respectively, and $\xor$ denotes the XOR operation. Nodes in Kademlia are solely identified by a single binary string. Hence, the common prefix length concern of \textit{\backupUpdate} is addressed by the XOR distance. Upon the locality-awareness of the Kademlia's identifiers, the scoring-based element replacement of \textit{\backupUpdate} also considers the latency constraint i.e., a lower XOR distance implies higher common prefix, and hence a lower underlying latency, which results in a higher score.

\begin{equation}
    e.score = \frac{e.sop}{e.id \xor id}
    \label{eq:interlace_update_scoring_kademlia}
\end{equation}

For \textit{\backupResolve} event handler, the node needs to change the $candCheck$ implementation (Algorithm \ref{intelace:alg_resolvingFailure}, Line \ref{intelace:alg_resolvingFailure:candCheck})  based on the architecture of DHT. For example, in Kademlia a (backup) neighbor is eligible to be contacted by the search protocol upon a lower XOR distance with the target identifier than the executor's one. Likewise, the DHT node needs to change the scoring mechanism of \textit{\backupResolve} event handler based on the identifiers' distance metric of the DHT overlay. In Kademlia's case that operates on XOR distances, the scoring mechanism needs to consider the XOR distance of each node to the search target as shown by Equation \ref{eq:interlace_resolve_scoring_kademlia}, which is the updated scoring formula of \textit{\backupResolve} where $target$ is the search target's identifier, and $commonPrefixLength$ is the length of common prefix between the identifiers of element $e$ and executor.

\begin{equation}
    e.score = \frac{e.sop \times commonPrefixLength}{e.id \xor target}
    \label{eq:interlace_resolve_scoring_kademlia}
\end{equation}

\section{\xspace \xspace Related Works}
\label{section:interlace_relatedworks}
\subsection{\textbf{Churn Stabilization}}
As a broad taxonomy, the P2P churn stabilization mechanisms for structured overlays are divided into static and dynamic approaches as described next.

\textbf{Static Churn Stabilization:} The aim of static approaches is to enrich the structural connectivity of the P2P overlay. D1HT \cite{monnerat2006d1ht}, Skip+ \cite{jacob2014skip+}, and EpiChord \cite{leong2006epichord} augment the overlay connectivity by providing \textit{larger routing tables}. These approaches need high maintenance overhead in terms of the number of update rounds and exchanged messages e.g., $O(\log^{2}{\systemCapacity})$ communication complexity for Skip+ \cite{jacob2014skip+}, which makes these approaches inefficient under continuous high churn rates where the overlay connectivity changes faster than the update rate of the nodes \cite{kaur2017churn}. As another example, D1HT needs $O(\systemCapacity)$ lookup table memory complexity, and epidemic dissemination of queries for the sake of maintenance. 
\textit{Hierarchical overlays}  \cite{kaur2016persistent, sacha2006discovery} cause unbalanced load on the upper tier nodes, and direct the P2P system towards centralization. Rainbow Skip Graph \cite{goodrich2006rainbow} and Tiara \cite{clouser2012tiara} follow the \textit{virtual nodes} approach where two or more physical peers are coupled into one single node of the P2P overlay to reduce the negative effect of individual departures on the overlay connectivity  \cite{meng2013using}. Virtual nodes, however, distort the structure of overlay and make it inappropriate for many applications including locality-aware replication \cite{hassanzadeh2016laras} that requires the nodes to be orchestrated based on their locality information \cite{hassanzadeh2015locality} rather than the availability.

\textbf{Dynamic Churn Stabilization:} In contrast to the static approaches that reinforce the structured overlay uniformly regardless of the churn, the dynamic churn stabilization approaches aim to maintain the connectivity based on the local knowledge of nodes about the underlying churn. The knowledge varies from a rigid assumption to intermittent perception from the underlying churn. The dynamic churn stabilization approaches are further classified into proactive, reactive, and predictive. In \textit{proactive} dynamic stabilization, nodes frequently check their neighbors' availability by either pinging them, or conducting a search for them \cite{rhea2004handling, medrano2015performance, li2004comparing, ghinita2006adaptive, zhao2004tapestry, paul2015interaction, li2005bandwidth}. The main disadvantage of proactive approaches is their dependency on active exploration of the overlay network, which leads a persistent communication overhead.

In \textit{reactive} churn stabilization approaches, the maintenance of routing tables is done only upon the detection of an entry's failure \cite{onana2003dks}. 
The threshold of failure at which the reactive stabilization starts is a function of the underlying churn rate. 
Reactive churn stabilization approaches are not bandwidth friendly. They congest the underlying network in the events where the churn rate is not well-predicted, or the threshold is not well-chosen. DKS \cite{onana2003dks} provides a structured P2P overlay construction that is in compliance with circular DHTs like Chord \cite{stoica2001chord} as well as those based on XOR distances e.g., Kademlia \cite{maymounkov2002kademlia}. To handle churn, each DKS node holds a fixed number of pointers to the nodes that immediately follow it in the identifier space. The list is updated upon failure detection of any of the successors in a "correction-on-used" manner i.e., the failed node is replaced by a new successor. DKS's success on maintaining connectivity is correlated with the failure pattern of the successors i.e., concurrent failiures of the successors keep a node away from finding new successors and recovering a search from failure. 1-backtracking \cite{kong2008resilience} is another reactive solution where a message is back-tracked one step upon detection of a failure on the path, and re-routed again. In systems with high churn rate or low availability, it is very likely for the alternative backtracked neighbors to be offline, which causes the entire search to be dropped but at a longer response time compared to no backtracking scenario. Also, increasing the degree of backtracking (e.g., 2-backtracking) results in high search delay due to the exponential growth of the alternatives that are contacted blindly without any availability perception. In contrast to \textit{\churnStabilization}, both DKS and backtracking approaches do not consider the search latency, search path length, and availability of the alternative neighbors for recovering a search message from failure. 

\textit{Predictive} churn stabilization is another dynamic approach where nodes aim to predict the failure of their neighbors ahead in time and redirect the search queries to the highly likely online neighbors. The general idea of predictive approaches is similar to our proposed \textit{\churnStabilization}. For example, Kademlia \cite{maymounkov2002kademlia} resembles \textit{\churnStabilization} in holding backup neighbors. However, in contrast to the \textit{\churnStabilization}, the backup neighbors in Kademlia are solely scored based on their last-seen time \cite{heck2017evaluating, li2004comparing, hojo2016frt}. 
Kademlia's approach on replacing the oldest entry with the newest piggybacked element increases the lack of routing candidates especially under high churn rates where the buckets are updated more frequently, which degenerates the connectivity of the system. Also, scoring the backup neighbors solely on a least-recently-seen basis increases the expected number of trials takes to find an online routing candidate, which increases the communication overhead as well as the query processing time, and exposes the underlying network to congestion in larger scales. Moreover, in contrast to \textit{\churnStabilization}, Kademlia does consider the latency and search path length in its scoring mechanism.

\subsection{\textbf{Availability Prediction}}
For availability prediction, \textit{regularity-based} solutions \cite{pace2011exploiting, song2009replica, kaur2017modeling} aim to extract the long-term regular behavior of nodes, and do not exploit irregular nodes, which constitute a large chunk of the system. Despite not showing a regular availability pattern, irregular nodes maybe online for a long enough while to participate in a churn stabilization protocol.
\textit{Vector-based} solutions \cite{lazaro2012long, hassanzadeh2016awake, ramachandran2012decentralized} predict the availability probability of the nodes within the slots of a fixed-periodic time interval e.g., availability probability in each hour of a day. \textit{Hidden Markov Model (HMM)} \cite{kaur2016performance} approaches predict the availability of a node qualitatively within the four states of born, young, aged, and offline. Both the vector-based and HMM-based approaches are long-term solutions that necessitate long learning phases, and are not applicable to instantaneous prediction cases like churn stabilization. Accordion \cite{li2005bandwidth} utilizes a variation of Lifetime predictor \cite{leonard2007lifetime}, where the availability probability of a node is computed as a function of its accumulative sessions' lengths. LUDP \cite{bustamante2008designing}  predicts the availability of nodes based on their number of incoming connections where a higher number of incoming connections as well as age corresponds to a higher availability probability. The age of a node is determined by its accumulative online sessions' lengths.

\subsection{\textbf{Algorithms used for comparison:}}
We selected Kademlia \cite{maymounkov2002kademlia} and DKS \cite{onana2003dks} for the sake of implementation and comparison with our proposed \textit{\churnStabilization} as they are the only ones that resemble \textit{\churnStabilization} in keeping the communication complexity of Skip Graph intact, and being needless of frequent probing. Among the availability prediction solutions, we selected DBG \cite{mickens2006exploiting}, Lifetime \cite{leonard2007lifetime}, and LUDP \cite{bustamante2008designing} for the sake of implementation and comparison with our proposed \textit{\swdbg} as they are the only ones that resemble \textit{\swdbg} in providing an instantaneous and fine-grained availability prediction. The implementation details of these algorithms are described next.
\subsubsection{Churn Stabilization}
\textit{Kademlia \cite{maymounkov2002kademlia}:} We follow the same implementation as \textit{\churnStabilization}, except, we replace the backup table's sets with double linked-lists. We distribute the backup size, $\backupSize$, uniformly among the buckets at every level and direction. 
In case the total number of levels is not a divisor of $\backupSize$, the remainder is equally distributed at each level by a value of two (i.e., one extra pointer at each direction) starting from level zero. The bottom-up distribution of the surplus backup is due to the importance of the bottom-most levels on the success of the search. Essentially, the level zero of Skip Graph is where an unsuccessful search terminates. Hence having a bigger backup size at this level increases the chance of the search being rescued from failure.  
We modify the \textit{\backupUpdate} to simply insert the piggybacked elements to the head of the proper linked-list, and remove linked-list's tail if the size goes beyond the permissible, $\backupSize$. We modify the \textit{\backupResolve} such that after identifying the proper bucket, it checks the online status of the routing candidates inside the bucket starting from the head until it finds an online routing candidate to return. Similar to \textit{\churnStabilization}, \textit{\backupResolve} returns NULL if the list runs out of routing candidate.

\textit{DKS \cite{onana2003dks}:} In our implementation of DKS, we distribute the backup size, $\backupSize$, among the levels similar to the Kademlia's case. Each Skip Graph node then continuously holds a list of pointers to its immediate neighbors at each level and direction. 
DKS does not piggyback any availability information on the search messages. Rather, upon joining the system, each node contacts its immediate neighbors and initializes its pointers list by invoking the \textit{\backupUpdate}. Also, in our implementation of the \textit{\backupResolve} a node selects the proper routing candidate from its pointer list that is identified by the level and direction of the search. If the selected routing candidate is offline, it is removed from the pointer list and the list's tail is updated by appending the immediate neighbor of the current tail. The \textit{\backupResolve} is invoked repeatedly in this manner until an online routing candidate is found, or no more immediate neighbor of the tail is available to be alternatively appended to the list of pointers. 

\subsubsection{Availability Prediction}
\textit{DBG \cite{mickens2006exploiting}}:
We implement DGB in similar way as \textit{\swdbg} with the current state window is being removed and only one DBG with fixed state size is employed. 

\textit{Lifetime \cite{leonard2007lifetime}}:  We compute the availability probability of each node at each time slot as the fraction of its accumulative session lengths to the total number of elapsed time slots since the beginning of the simulation. 

\textit{LUDP \cite{bustamante2008designing}}: We quantified the availability probability of each node as a function of its age (i.e., the overall number of time slots it has been online) as well as its total number of incoming connections as shown by Equation \ref{eq:interlace_ludp} where $op_{t}$ is the online probability of the node at the $t^{th}$ time slot, $T_{o}$ is the age of the node, and $Num_{in}$ denotes the total number of incoming connections of the node.

\begin{equation}
    op_{t} = \frac{T_{o} \times Num_{in}}{t \times  \systemCapacity}
    \label{eq:interlace_ludp}
\end{equation}

\section{\xspace \xspace Simulation Setup}
\label{section:interlace_simulation}
To simulate and evaluate the churn stabilization solutions, we extended the Skip Graph simulator SkipSim \cite{skipsim} by enabling arrival, searching, and departure under the crash-failure model where nodes depart the system without notifying their overlay neighbors. Among the existing churn models of P2P systems, 
we found the BitTorrent-based models in \cite{stutzbach2006understanding} stronger, more realistic, reliable, parametrically clearer than the others. Therefore, we applied the same Debian churn distribution as \cite{stutzbach2006understanding} on SkipSim. Debian churn model follows a Weibull distribution with the average session length of $2.71$ hours, and the average interarrival time of $39.86$ seconds. We implemented \textit{\churnStabilization}, \textit{\swdbg} as well as the best existing churn stabilization and availability prediction approaches that are applicable to our system model, and simulated all approaches under the specified Debian churn model.
The average downtime of the nodes (i.e., average offline time) is correlated with the size of the system. We provide analysis over downtime in Section \ref{section:interlace_results}. 


We consider the time as discrete with fixed time slots of one hour. The Skip Graph is initially empty of nodes. At the beginning of each time slot, some nodes arrive the system, join the Skip Graph overlay, and start interacting with others via the search for numerical ID protocol. Such interaction is the basic operation of our system model that is required for example by the replication \cite{hassanzadeh2018decentralized, hassanzadeh2016laras, hassanzadeh2016awake}, or aggregation \cite{hassanzadeh2017elats} protocols. The number of arrivals to the system is determined by the interarrival time distribution. 
Likewise, in each time slot, SkipSim selects the number initiated searches uniformly between $[0, {\systemCapacity_{o} \choose 2}]$ where $\systemCapacity_{o}$ denotes the number of online nodes in that time slot. The initiator and target numerical IDs of each search are uniformly selected from the set of online nodes in that time slot. Each node is only available for a limited number of time slots that follows the session length distribution of the churn model and leaves the system once its session length is over. In SkipSim a node departs the system at the end of the time slot that its session length terminates, and becomes offline. 

We simulated each algorithm for $100$ randomly generated topologies, each with the system capacity of $1024$ nodes. Each topology was simulated for one week (i.e., $168$ time slots). In order to simulate pinging, each time during a search a node checks the online status of another node, SkipSim adds the corresponding round trip time (RTT) to the total search time.

\section{\xspace \xspace \xspace Analytical and Performance Results}
\label{section:interlace_results}
\subsection{Analytical Framework}
We present a framework that analyzes the success probability of \textit{\churnStabilization} as a function of the backup size $\backupSize$ on routing search messages under uniform churn model. In the uniform churn model, at any time, the probability that a given node goes offline is a constant denoted by $\uniformFailureProb$, which is independent of the failure of other nodes. With this framework, we provide a conservative estimation of backup size to achieve the maximum success ratio of searches, and support it with experimental results that are presented in Section \ref{section:interlace_results}. 
In our framework, we assume a P2P system of size $\systemCapacity$ i.e., $\systemCapacity$ nodes with unique numerical IDs from $1$ to $\systemCapacity$. For each node, we model the sample space for choosing a random neighbor as the set of all piggybacked identifiers of other nodes that the node receives upon routing search requests. Moreover, to generalize the sample space for every chosen node, we assume that the node routes enough search requests for the sample space size to converge to $\systemCapacity$. Let $X$ be a random variable denoting a uniformly chosen numerical ID from the set $[1,\systemCapacity]$. The probability of $X$ has a certain numerical ID of $x$ is denoted by Equation \ref{interlace:eq_pr_x}.

\begin{equation}
    Pr(X = x) = \frac{1}{\systemCapacity} \label{interlace:eq_pr_x}
\end{equation}

Let $T$ be another random variable denoting a uniformly chosen search for numerical ID target from the set of numerical IDs of all the nodes in system. Once node $x$ is determined, the probability that a search for a randomly chosen numerical ID of $t$ in the right direction reaches $x$ is denoted by Equation \ref{interlace:eq_pr_t_given_x}. Right direction of the search corresponds to $x \leq t$ in the conditional probability of Equation \ref{interlace:eq_pr_t_given_x}. The probability is taken over the set of all numerical IDs that are greater than $x$ and less than $t$. The right search direction is assumed without loss of generality, and a similar analysis is applicable to the left direction.

\begin{equation}
        Pr(T = t \land x \leq t | X = x) = \frac{1}{\systemCapacity-x+1} \label{interlace:eq_pr_t_given_x}
\end{equation}

We start by the simplest scenario where a node $x$ holds only one right neighbor address. Let $Y$ be another random variable that corresponds to the numerical ID of the sole right neighbor of node $x$. We define $p_{x,t}$ that is shown by Equation \ref{interlace:eq_pr_y} as the probability of node $Y$ taking a numerical ID value in $(x,t]$, and hence being a proper routing candidate for node $x$ to forward the search for numerical ID of $t$ to it. In other words, $p_{x,t}$ is the probability of the search successfully passing $x$ and proceeding on right direction towards $t$ via node $Y$.

\begin{equation}
     p_{x,t} = Pr(x \leq Y \leq t | X = x, T = t, x \leq t) = \frac{t - x}{\systemCapacity} \label{interlace:eq_pr_y}    
\end{equation}

We define $p$ as the marginal probability of $p_{x,t}$ with respect to both $x$ and $t$ i.e., the probability of a uniformly chosen node as the right neighbor of a node on a search path for a target numerical ID being a proper routing candidate. By being a routing candidate, we mean that the numerical ID of the chosen neighbor lays between the node's and the target's numerical IDs. As shown by Equations \ref{interlace:eq_p_1}-\ref{interlace:eq_p_3}, $p$ is directly derived by taking the sum of products of Equations \ref{interlace:eq_pr_x}, \ref{interlace:eq_pr_t_given_x}, and \ref{interlace:eq_pr_y} over all possible values of $x$ and $t$. As $x$ moves over the numerical ID domain of $[1,\systemCapacity]$, $t$ varies between $[x,\systemCapacity]$, which implies that the search target $t$ should be greater than or equal to $x$.

\begin{align}
     p &= \sum_{x = 0}^{\systemCapacity} \sum_{t=x}^{\systemCapacity} p_{x,t} \times Pr(T = t| X = x, x \leq t) \times Pr(X = x) \label{interlace:eq_p_1}\\
       &= \sum_{x = 0}^{\systemCapacity} \sum_{t=x}^{\systemCapacity} \frac{t - x}{\systemCapacity} \times \frac{1}{\systemCapacity-x+1} \times \frac{1}{\systemCapacity} \label{interlace:eq_p_2}\\ 
       &= \frac{1}{\systemCapacity^2}\sum_{x = 0}^{\systemCapacity} \sum_{t=x}^{\systemCapacity} \frac{t - x}{\systemCapacity-x+1}
       \label{interlace:eq_p_3}
\end{align}

We skip the intermediate computations of summation for the sake of space, which yields that the probability $p$ converges to $\frac{1}{4}$ as the system size approaches infinity. This conveys that each node of the system choosing only one of the other nodes uniformly as its right neighbor, the probability that the chosen right neighbor is a routing candidate for a randomized search target is about $\frac{1}{4}$. 
Assuming the uniform churn model with the failure probability of $\uniformFailureProb$
turns the probability $p$ to the more realistic $p' = p \times (1 - q) \approx \frac{1 - q}{4}$ i.e., the probability in which a uniformly chosen right neighbor for a uniformly chosen node on a search path for a uniformly chosen target is an online routing candidate. 

Let random variable $Z$ denotes the number of online routing candidates for a uniformly chosen node on the search path of a uniformly chosen numerical ID among the $\backupSize$ many backup neighbors that are chosen uniformly. The uniformly chosen backup neighbors is the weakened version of \textit{\churnStabilization} that we utilize as a conservative analytical baseline. $Z$ is a Binomial random variable with success probability of $p'$ \cite{bertsekas2002introduction}. We define the failure probability $p_{f}$ as the probability of having no online routing candidate for a node on a search path of a search for numerical ID as shown by Equation \ref{interlace:eq_p_f}.

\begin{equation}
    p_{f} = \binom{\backupSize}{0} \times (p')^0 \times (1-p')^\backupSize = (1- p')^\backupSize \label{interlace:eq_p_f}
\end{equation} 

Treating $p_{f}$ as a Bernoulli probability, we denote the expected path length of a search for a uniformly chosen numerical ID that leads to failure with $E_{f}$ as shown by Equation \ref{interlace:eq_e_f}. In other words, $E_{f}$ corresponds to the expected number of nodes that a search message should traverse to face a failure with high probability.

\begin{equation}
   E_{f} = \frac{1}{p_{f}} = \frac{1}{(1- p')^\backupSize} = \frac{1}{(1-\frac{1-q}{4})^\backupSize} \label{interlace:eq_e_f}
\end{equation}

Considering the failure probability of nodes (i.e., $q$) as a system-wide constant bound, $E_{f}$ is tweakable with the size of backup table i.e., $\backupSize$. Equation \ref{interlace:eq_e_f} implies that in order to achieve no failure in expectation with high probability, $\backupSize$ should be chosen large enough that $E_{f}$ stays beyond the average search path length of the system.

For example, in the specified Debian churn model (see Section \ref{section:interlace_simulation}), each node has an average session length of about $2.71$ hours and then goes offline. Having an average inter-arrival time of $39.86$ seconds results in an average of $90$ hourly arrivals, which causes an offline node to return back to the system in about $12$ hours in expectation. Modeling this behavior with a uniform churn model results in a uniform failure probability of about $0.82$ that is analogous to $q$ in our proposed framework. In a system with $\systemCapacity = 1024$ nodes under this uniform churn model, the expected number of online nodes at each time slot is about $184$ that is obtained from Equation \ref{interlace:eq_online_nodes}, and is denoted by $E_{online}$. During the simulation, we consider a node as online within a (one hour) time slot, if it arrives at the system at that time slot.

\begin{equation}
    E_{online} = (1 - q) \times \systemCapacity = 0.18 \times 1024 \approx 184 \label{interlace:eq_online_nodes}
\end{equation}

\noindent The search path length in Skip Graph is asymptotically logarithmic in the number of nodes i.e., a Skip Graph with $\systemCapacity$ nodes experiences search paths' length of $O(\log{\systemCapacity})$ nodes. Having an average of about $184$ online nodes in the Skip Graph at every time slot, we approximate the lower bound on the average search path length as $\ceil{\log{184}} = 8$ nodes. Assuming $E_{f} = 8$ and $q = 0.82$, we obtain $b \approx 40$ from Equation \ref{interlace:eq_e_f} as an estimate on the backup size that maximizes the average success ratio of searches under the specified Debian churn model. 

\subsection{Performance Results} 
\textbf{Availability Prediction:} 
Table \ref{interlace:table_availability_prediction_error} represents a comparison between the average prediction error of our proposed \textit{\swdbg} and the existing availability prediction solutions. 
For the sake of comparison, we measure the availability prediction error as the average difference between the nodes' predicted availability probability, and their availability status. When a node is offline, its availability status is equal to $0$, and when it becomes online, its availability status is $1$. The average is taken over all the simulation's time slots. 
DBG($x$) denotes the DBG implementation with the state size of $x$. Compared to DBG($4$) that performs as the best existing availability predictor, \textbf{\textit{\swdbg} predicts the availability of nodes with about $\predictionGain$ times more accuracy} under the Debian churn model. Growing the state size of DBGs beyond $4$-bits increases their state update running time exponentially in their state size, and is not applicable to our simulation scale.   \\

\begin{table*}
\centering
\resizebox{\columnwidth}{!}
{
    \begin{tabular}{ |l|l|l|l|l|l|l|l| }
    \hline
    Prediction Strategy & \textbf{\textit{\swdbg}} & DBG($1$) & DBG($2$) & DBG($3$) & DBG($4$) & LUDP & Lifetime \\
    \hline
    Average Prediction Error & \textbf{0.18} & 0.28 & 0.26 & 0.23 & 0.21 & 0.51 & 0.34 \\
    \hline
    \end{tabular}
}
\caption{Average prediction error of availability prediction strategies under Debian churn model.}
\label{interlace:table_availability_prediction_error}
\end{table*}

\textbf{Average Success Ratio: }
Consistent with existing churn stabilization studies like \cite{herrera2007modeling}, we consider the average success ratio of searches as the connectivity performance of Skip Graph overlay under churn. Figure \ref{interlace:fig_performance}.a shows connectivity performance of churn stabilization approaches in the specified Debian churn model as the backup size (i.e., the $\backupSize$ parameter) increases. 
\textit{\churnStabilization}-$x$ represents the setup with \textit{\churnStabilization} as the churn stabilization approach, and the availability prediction strategy denoted by $x$.   
As illustrated in Figure \ref{interlace:fig_performance}.a, there is a similar connectivity performance pattern among almost all the \textit{\churnStabilization}-$x$ approaches i.e., the success ratios start by a fast growth for smaller backup sizes up to a common breakpoint in which the growth rate slows down and gets steady. For example, \textit{\churnStabilization}-\textit{\swdbg} experiences a drastic increase of success ratio in moving from the backup size of $10$ to $20$ (nodes) followed by a narrow increase after passing the breakpoint of $20$, and converges to a steady state value of success ratio of about $0.9$. As detailed earlier, we obtain $b \approx 40$ from our analytical framework as an estimate on the backup size that maximizes the average success ratio of searches under the specified Debian churn model. As shown in Figure \ref{interlace:fig_performance}.a, for almost all of the \textit{\churnStabilization}-$x$ approaches, backup size of $40$ yields a success ratio in the steady-state part of the graph that converges to the maximum value. $b = 40$ is a conservative estimate on backup size, and similar connectivity performance is obtainable with even lower backup sizes (e.g., $b = 20$). However, it is yet a proper estimator to maximize the connectivity performance of the \textit{\churnStabilization}-$x$ approaches under the characterized churn model.

\begin{figure*}
\centering
\begin{minipage}[c]{.45\textwidth}
\centering
    \scalebox{0.6}{\begin{tikzpicture}
\begin{axis}[
    ylabel={Average Success Ratio},
    xlabel={Backup Size (nodes)},
    xmin=0, xmax=50,
    ymin=0, ymax=1,
    xtick={0, 10, 20, 30, 40, 50},
    ytick={0, 0.1, 0.2, 0.3, 0.4, 0.5, 0.6, 0.7, 0.8, 0.9, 1},
    legend style={
    legend pos=outer north east, legend cell align = left},  
    ymajorgrids=true,
    grid style=dashed,
]

    \addplot[mark size = 4,color = red,  mark = star, mark color = red]
    coordinates 
    {
        (10,  .72)
        (20,  .81)
        (30,  .83)
        (40,  .83)
        (50,  .85)
    };
    
    \addplot[mark size = 4,color = orange,  mark = +, mark color = orange]
    coordinates 
    {
        (10,  .53)
        (20,  .58)
        (30,  .59)
        (40,  .60)
        (50,  .60)
    };

    \addplot[mark size = 4,color = blue,   mark = +, mark color = blue]
    coordinates 
    {
        (10,  .61)
        (20,  .68)
        (30,  .69)
        (40,  .70)
        (50,  .70)
    };
    
    \addplot[mark size = 4,color = purple,   mark = +, mark color = purple]
    coordinates 
    {
        (10,  .64)
        (20,  .71)
        (30,  .72)
        (40,  .73)
        (50,  .73)
    };
    
    \addplot[mark size = 4,color = black,   mark = +, mark color = orange]
    coordinates 
    {
        (10,  .66)
        (20,  .73)
        (30,  .74)
        (40,  .75)
        (50,  .75)
    };
    
    \addplot[mark size = 4,color = blue,   mark = square, mark color = blue]
    coordinates
    {
        (10,  .24)
        (20,  .24)
        (30,  .24)
        (40,  .24)
        (50,  .24)
    };

    \addplot[mark size = 4,color = black,   mark = square, mark color = black]
    coordinates
    {
        (10,  .30)
        (20,  .37)
        (30,  .39)
        (40,  .39)
        (50,  .39)
    };
    
    \addplot[mark size = 4,color = orange,   mark = square, mark color = orange]
    coordinates 
    {
        (10,  .30)
        (20,  .42)
        (30,  .49)
        (40,  .53)
        (50,  .56)

    };

    \addplot[mark size = 4,color = green,   mark = square, mark color = green]
    coordinates
    {
        (10,  .22)
        (20,  .23)
        (30,  .24)
        (40,  .25)
        (50,  .25)
    };
    

\end{axis}
\end{tikzpicture}}
    \caption*{(a)}
\end{minipage}
\begin{minipage}[c]{.45\textwidth}
\centering
    \scalebox{0.6}{\begin{tikzpicture}
\begin{axis}[
    ylabel={Average Search Latency (s)},
    xlabel={Average Success Ratio},
    xmin=0, xmax=1,
    ymin=0, ymax=60,
    xtick={0, 0.1, 0.2, 0.3, 0.4, 0.5, 0.6, 0.7, 0.8, 0.9, 1},
    ytick={0,5,10,15,20,25,30,35,40,45,50, 55, 60},
    legend style={
    legend pos=outer north east, legend cell align = left},  
    ymajorgrids=true,
    grid style=dashed,
]

    \addplot[color = red, only marks, mark size = 4, mark = star, mark color = red]
    coordinates 
    {
        (.72,  20.33)
        (.81,  14.07)
        (.83,  12.85)
        (.84,  12.51)
        (.85,  12.17)
    };
    
    \addplot[color = orange, only marks, mark size = 4, mark = +, mark color = orange]
    coordinates 
    {
        (.53,  17.49)
        (.58,  14.47)
        (.59,  13.93)
        (.60,  12.97)
    };

    \addplot[color = blue,  only marks, mark size = 4, mark = +, mark color = blue]
    coordinates 
    {
        (.61,  19.23)
        (.68,  15.21)
        (.69,  13.97)
        (.70,  13.00)
    };
    
    \addplot[color = purple,  only marks, mark size = 4, mark = +, mark color = purple]
    coordinates 
    {
        (.64, 19.70)
        (.71, 15.06)
        (.72, 13.75)
        (.73, 13.10)
    };
    
    \addplot[color = black,  only marks, mark size = 4, mark = +, mark color = orange]
    coordinates 
    {
        (.66, 19.98)
        (.73, 14.92)
        (.74, 13.68)
        (.75, 13.09)
    };
    
    \addplot[color = blue,  only marks, mark size = 4, mark = square, mark color = blue]
    coordinates
    {
        (.24, 14.19)
    };
    
    \addplot[color = black,  only marks, mark size = 4, mark = square, mark color = black]
    coordinates
    {
        (.35,  15.87)
        (.37,  12.86)
    };
    
    \addplot[color = orange,  only marks, mark size = 4, mark = square, mark color = orange]
    coordinates 
    {
        (.30, 23.76)
        (.42, 30.52)
        (.49, 36.20)
        (.53, 41.54)
        (.56, 45.94)

    };
    
    \addplot[color = green,  only marks, mark size = 4, mark = square, mark color = green]
    coordinates
    {
        (.22, 12.30)
        (.23, 33.04)
        (.24, 49.60)
        (.25, 55.77)
    };
    
    \legend{
    \churnStabilization-\swdbg,
    \churnStabilization-DBG-1, \churnStabilization-DBG-2, \churnStabilization-DBG-3, \churnStabilization-DBG-4,  \churnStabilization-LUDP,
    \churnStabilization-Lifetime,
    Kademlia, DKS}

\end{axis}
\end{tikzpicture}}
    \caption*{(b)}
\end{minipage}
\caption{(a) Average success ratio of searches vs the backup size. X-axis corresponds to the backup size (i.e., $\backupSize$), and Y-axis corresponds to the average success ratio of the searches. The desired data points are the ones towards the top right corner, which correspond to a higher connectivity of overlay under churn, compared to the rest. (b) Average search latency vs average success ratio of the searches. X-axis corresponds to the average success ratio, and Y-axis corresponds to the average search latency in seconds. The desired data points are the ones towards the bottom right corner, which correspond to a higher connectivity of overlay under churn at a lower latency, compared to the rest.}
\label{interlace:fig_performance}
\end{figure*}

The scoring strategy of \textit{\churnStabilization} on the backup neighbors is directly affected by the underlying availability prediction accuracy. Supported by Table \ref{interlace:table_availability_prediction_error}, the prediction error of DBGs narrows down as their state size increases, which results in a direct relationship between the average success ratio of \textit{\churnStabilization}-DBG approaches and the state size of DBG. \textit{\churnStabilization}-Lifetime follows the breakpoint pattern but at a lower success ratio rate compared to the DBG-based approaches. This follows from the Lifetime prediction mechanism's inferiority compared to the DBG-based approaches in predicting the availability of nodes. \textit{\churnStabilization}-LUDP is the only \textit{\churnStabilization}-$x$ approach that does not follow the breakpoint pattern. This is due to the $0.51$ prediction error margin of LUDP, which provides a completely randomized prediction of availability, and results in \textit{\churnStabilization}-LUDP to perform as the weakest compared to the rest. DKS acts closely to LUDP as the second weakest, which is due to failure on recovering from the concurrent departures of consecutive neighbors. Higher backup sizes increase the success ratio of Kademlia with more routing candidates provided. However, the coarse-grained approach of Kademlia on replacing the oldest entry with the new candidate as well as keeping fixed size backup list at each level results in the lack of routing candidates within proximity of the nodes in the identifier space. This makes Kademlia perform less efficiently in the lower backup sizes, and converge to \textit{\churnStabilization}-DBG($1$) in the higher backup sizes.  Benefiting from \textit{\swdbg} that adaptively chooses the best state size for each node, as well as scoring the backup table elements based on both their numerical ID distances and availability behavior features \textit{\churnStabilization}-\textit{\swdbg} the best among the others. 
Compared to the Kademlia that acts as the best existing solution applicable on Skip Graph, \textbf{\textit{\churnStabilization}-\textit{\swdbg} improves the average success ratio of the searches with the gain of about $\successRatioGain$} times on average. \\

\textbf{Average Search Latency: }
Figure \ref{interlace:fig_performance}.b shows the the average search latency of Skip Graph under churn versus the average success ratio of the search for different churn stabilization approaches. A point $(x,y)$ on this figure is interpreted as \textit{$y$ is the best average search latency that is provided by the corresponding churn stabilization approach to maintain the average search success ratio of $x$}. The desired data points in Figure \ref{interlace:fig_performance}.b are the ones towards the bottom right corner, which correspond to a higher connectivity of overlay under churn at a lower latency, compared to the rest.
There exists a correlation between the prediction error of the availability predictors (Table \ref{interlace:table_availability_prediction_error}) and the average search latency (Figure \ref{interlace:fig_performance}.b).
A higher accuracy of availability prediction yields in maintaining a higher number of likely available backup neighbors. 
This lowers the expected number of timeouts that contribute to the overall search latency, which reduces the average search latency. 
Ignoring the availability of neighbors in DKS significantly increases its search latency as the backup size grows (i.e., larger success rations of search). A similar pattern occurs for Kademlia with loosened availability constraint on the selection of routing candidates.
Affected by the immense availability prediction error of LUDP, the \textit{\churnStabilization}-LUDP approach initially keeps likely unavailable backup neighbors that are later dropped by \textit{\churnStabilization} due to their timeout failures. Therefore, as the time goes on, the majority of nodes are with depleted back up tables, which conclude the search faster but with the wrong result. This is why \textit{\churnStabilization}-LUDP seems faster than the rest but at a miserably lower average success ratio of the search.
Similar to the connectivity performance, our proposed \textbf{\textit{\churnStabilization}-\textit{\swdbg} outperforms Kademlia by performing the searches $\successRatioGain$ times more successful, and $\searchTimeGain$ times faster on average, and acts as the best among all the existing counterparts.} 

\subsection{Space, Time, and Communication Complexities}
\textit{\swdbg}: As DBGs are sorted in ascending order based on their state size in the current state window, the size of \textit{\swdbg} is bound by the Right DBG's size. Having the state size of $k$ for the Right DBG in the current state window, the asymptotic time and memory complexity of \textit{statusUpdate} (Algorithm \ref{intelace:alg_onlinestateUpdate}) is $O(2^{k})$. Despite this exponential asymptotic upper bound, based on our simulation, the average state size of the Right DBG is about $3.6$ bits and does not cross the $5$ bits. The standard deviation of average Right DBG state size of \textit{\swdbg} is about $0.2$.  
Representing each vertex by a $k$-bit state string and an integer probability value between $0$ and $100$ applies an average and maximum memory overhead of about $64$ and $256$ bytes on each node that utilizes \textit{\swdbg}. 
Running on Intel i5 $2.60$ GHz CPU and $8$ GB of RAM, a single execution of \textit{statusUpdate} takes the average running time of $1.45$ milliseconds. \textit{\swdbg} does not impose any communication overhead on the system.

\textit{\churnStabilization}:
The only memory overhead of \textit{\churnStabilization} on a node is a backup table size of $O(\backupSize)$ where $\backupSize$ is a system wide constant that denotes the maximum backup size. 
As each of the \textit{\churnStabilization} event handlers iterates over the backup table either partially (i.e., one entry set) or entirely, the worst-case asymptotic running time of \textit{\churnStabilization} is  $O(\backupSize)$.
 The worst-case communication complexity of \textit{\backupUpdate} and \textit{\backupResolve} are $O(1)$ and $O(\backupSize)$, respectively. The asymptotic $O(\backupSize)$ worst case communication complexity of \textit{\backupResolve} occurs when all the backup neighbors are placed in a single entry set, and are contacted one by one on resolving a failure. However, the expected number of backup neighbors for each level of a node is $\frac{\backupSize}{\log{\systemCapacity}}$, which applies the expected communication complexity of $O(\frac{\backupSize}{\log{\systemCapacity}})$. As long as $b = O(\log^{2}{\systemCapacity})$, \textit{\churnStabilization} does not change the communication complexity of the Skip Graph overlay. 
Based on the simulation results, for the backup size of $\backupSize = 50$, the average number of backup neighbors for each level of Skip Graph remains close to its expected value (i.e., an average of about $3.32$), that imposes the average communication complexity of $1.55$ messages per invocation of \textit{\backupResolve} by each node on a search path.

\section{Conclusion}
\label{section:interlace_conclusion}
To maximize the connectivity of Skip Graph-based DHT overlays under churn we proposed \textit{\churnStabilization}, a fully decentralized predictive churn stabilization algorithm that provides fine-grained scoring mechanisms based on the distribution of nodes in both the overlay and identifier space, as well as their availability probability. \textit{\churnStabilization} does not change the asymptotic communication complexity.
As an independent contribution, we proposed \textit{\swdbg}, a tool to predict the availability probability of the nodes. 

We extended the Skip Graph simulator, SkipSim \cite{skipsim}, implemented and simulated the state-of-the-art availability prediction methods as well as churn stabilization approaches that are applicable on a Skip Graph overlay. Our simulation results show that compared to the best existing solutions that are applicable on a Skip Graph overlay, \textit{\churnStabilization} improves the connectivity of the Skip Graph overlay under churn with the gain of about \textbf{$\successRatioGain$} times. Likewise, compared to the existing availability prediction approaches for P2P systems, \textit{\swdbg} is about \textbf{$\predictionGain$} times more accurate. A Skip Graph that benefits from \textit{\churnStabilization} and \textit{\swdbg} is about $\searchTimeGain$ times faster on average in processing the search queries under churn compared to the best existing solutions. 
\section*{Acknowledgement}
The authors thank
Muharrem Salel for his contributions to the SkipSim implementation.

\bibliographystyle{IEEEtran}
\bibliography{references}

\end{document}